\documentclass[twocolumn,pre,amsmath,amssymb]{revtex4-1}
\usepackage{graphicx}
\DeclareMathOperator*{\arcsec}{arcsec}

\begin{document}

\title{Solution of the Roth-Marques-Durian Rotational Abrasion Model}
\author{Bryan Gin--ge Chen}
\affiliation{Department of Physics and Astronomy, University of
Pennsylvania, Philadelphia, PA 19104-6396}
\date{\today}

\begin{abstract}
We solve the rotational abrasion model of Roth, Marques and Durian
[Phys.~Rev.~E (2010)],
a one-dimensional quasilinear partial differential equation resembling
the inviscid Burgers equation with the unusual feature of a step
function factor as a coefficient.  The complexity of the solution is
primarily in keeping track of the cases in the piecewise function that 
results from certain amputation and interpolation processes, so we 
also extract from it a model of an evolving planar tree graph that 
tracks the evolution of the coarse features of the contour.
\end{abstract}

\maketitle

What determines the shapes of pebbles is an intriguing 
physical question with interest not just to beachcombers out for walks 
but also geologists, who are interested in the history of
erosion at a site \cite{Boggs01}, as well as mechanical engineers 
\cite{Rabin95}, who wish to understand wear processes.  Recently
several models have been proposed to explain these shapes.  Two stochastic 
models are of note, a ``cutting model'' \cite{Durian06,Durian07}
which accompanied an experimental measurement of pebbles rotating in a tray 
and an analytically tractable ``chipping model'' \cite{Krapivsky07}.
These models lead to distributions of non-circular shapes.  
More recently, deterministic erosion processes have been studied by Roth, 
Marques and Durian.  They performed an experiment to measure 
the contours of linoleum tiles of fixed thickness and varying shape that they had 
rotated for differing amounts of time in a slurry of grit \cite{RMD10}.  
This paper describes the solution to their rotational abrasion
model.  Supposing $r(\theta,t)$ describes the radial 
distance of the contour from the rotational axis as a function of angle 
$\theta$ and time $t$, they proposed
\begin{align}\label{eqn:ev0}
\frac{\partial r}{\partial t}+Cr
\frac{\partial r}{\partial \theta}H\left(\frac{\partial
r}{\partial\theta}\right)&=0.
\end{align}
Here $C$ is a positive proportionality factor with
dimensions of Angle/(Time$\times$Length) and $H(x)$ denotes a
Heaviside step function defined so that $H(x)=0$ for $x\leq0$ and
$H(x)=1$ for $x>0$.  It is obvious that circles
($r(\theta,t)=$ constant) are stationary solutions to this equation,
and the evolution of experimental contours computed in \cite{RMD10} by
a finite differencing scheme also evolved towards circles unerringly.
These solutions also matched quantitatively the evolution of 
several geometrical quantities extracted from their experimental data, 
such as area, perimeter, and the width of the curvature distribution.

Summarized here are the key ideas behind our exact solution 
of this equation.  First, we exploit a connection to the Burgers 
equation at zero viscosity, a well-studied equation from gas
dynamics \cite{Whitham}.  Second, the solution $r(\theta,t)$ can be written 
in a piecewise 
fashion as a union of $r=$ constant (circular) arcs and certain 
``stretched'' segments of the initial contour.  More precisely, 
these segments are curves of the 
form $r_0(\theta(\theta_0,t))$ where $r_0(\theta_0)$ is the initial
contour and $\theta(\theta_0,t)$ at fixed $\theta_0$ is a linear 
function in $t$.  
The solution is constructed to be continuous, but will admit corners
with discontinuous slope generically.  Finally, the organization of
the solution has a strong combinatorial flavor, and the evolution of the 
pattern of critical points in the contour is captured by a model of an 
evolving planar tree.  This reduction suggests that discrete
statistical models may capture the properties of ensembles of abraded 
pebbles.  The final section discusses other possible extensions.

\section{Method of characteristics}

In what follows, we will usually think of the variables $r$ and
$\theta$ in Eq.~(\ref{eqn:ev0}) as two dimensional rectangular coordinates 
(indeed, all the figures are plotted in this ``unwrapped''
fashion), though they do refer to polar coordinates, and we will usually refer 
to curves of constant $r$ as circular arcs and to lengths in the $\theta$ 
direction as angular widths. 

We first observe that Eq.~(\ref{eqn:ev0}) without
the step function factor is precisely the inviscid Burgers equation, a
quasilinear first-order partial differential equation.
That equation may be solved by the method of
characteristics, which we shall now adapt.  See also the book of
Melikyan on solutions via characteristics to nonsmooth first-order 
equations in the theory of optimal control and in differential 
games \cite{Melik98}.  Let the initial contour be
$r_0(\theta)\equiv r(\theta,0)>0$.  We now search for 
``characteristics'', or space-time curves $\theta(\sigma),t(\sigma)$ 
beginning at $\theta(0)=\theta_0$, $t(0)=0$  ($\sigma$ being some 
parameter) 
along which $r(\theta(\sigma),t(\sigma))$ remains constant,
and hence equal to $r_0(\theta_0)$.  In other words, each point on the
initial contour $r_0(\theta_0)$ evolves forward in time along
its characteristic.

By applying the chain rule, we find
\begin{align*}
\frac{d}{d\sigma}r(\theta(\sigma),t(\sigma))&=0\\
\frac{d\theta}{d\sigma}\frac{\partial r}{\partial
\theta}+\frac{dt}{d\sigma}\frac{\partial r}{\partial t}&=0.
\end{align*}

Comparing this to Eq.~(\ref{eqn:ev0}), 
\begin{align*}
\frac{dt}{d\sigma}&=1\\
\frac{d\theta}{d\sigma}&=CrH\left(\frac{\partial r}{\partial
\theta}\right).
\end{align*}

Integrating these equations with the initial condition $t=0$, 
$\theta=\theta_0$ and using the fact that $r(\theta(\sigma),t(\sigma))$ 
is constant and equal to $r_0(\theta_0)$, we obtain
\begin{align}\label{eqn:char}
r(\theta(\theta_0,t),t)&=r_0(\theta_0)\\
\theta(\theta_0,t)&=\theta_0+Cr_0(\theta_0)tH\left(\frac{\partial r}{\partial
\theta}\right).
\end{align}

\begin{figure}
\centering
\includegraphics[width=7cm]{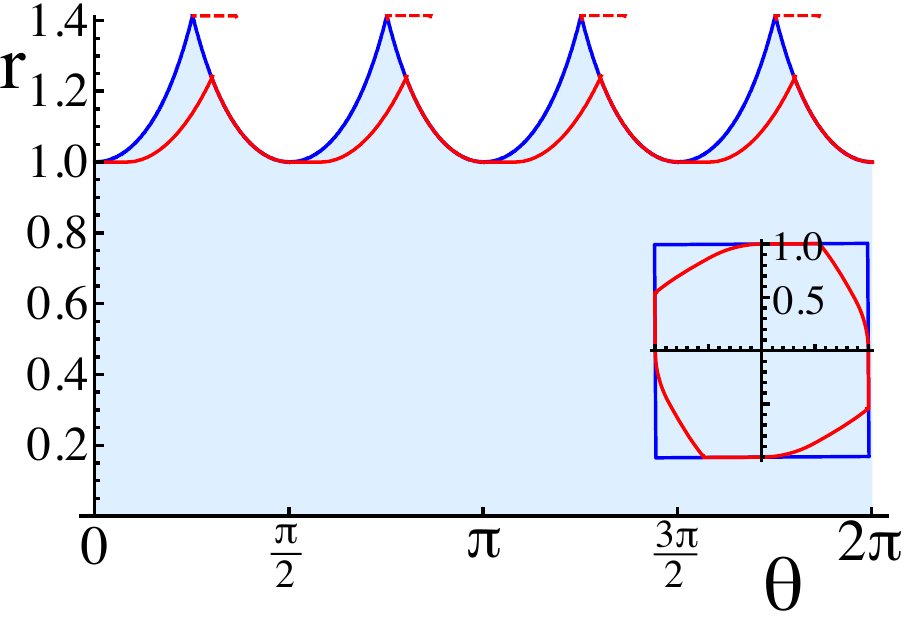}
\caption{(Color online) Evolution of a square, $r$ vs $\theta$ ($C$=1) via 
the method of characteristics. $t=0$ (blue) and $t=0.04$ (red). Dotted 
portions of the curve indicate pieces of the evolved contour which
are amputated, and the horizontal dashes indicate discontinuities of
the curve before amputation. The inset shows the two contours plotted in 
polar coordinates.}
\label{fig:frictionev}
\end{figure}

This equation yields a (possibly multi-valued) formal solution for 
$r(\theta,t)$ via
$r(\theta(\theta_0,t),t)$.  To plot the evolution of a curve 
with the solution in this form, begin with points distributed on the 
initial curve and move each of those points along its 
characteristic arc an angular distance 
$d\theta=Cr_0(\theta_0)H(\partial_\theta r)dt$ in each time step 
$dt$.  See Fig.~\ref{fig:frictionev} showing this evolution in the
case of an initially square contour.  

% define a hump?
\section{Multivaluedness and amputation}

The solution given in the previous section is not yet well-defined; 
we must deal with the 
multivaluedness of the evolution along characteristics.  To see how 
this arises, consider Fig.~\ref{fig:frictionev}.  
Points on the contour slightly behind the local 
maxima will quickly overtake the points with the same 
$r$-values but slightly ahead of the maximum, as those have
nonpositive $\partial_\theta r$ and hence are frozen by the step
function factor.  This causes the two pieces of the contour to overlap, 
and is analogous to the formation of shocks in the inviscid Burgers 
equation.  The ``horizontal'' discontinuity this creates is depicted in 
the figure as dotted lines.  Based on the physical interpretation of
the equation, the way to deal with this
multivaluedness is to amputate the portion of the contour where this
has occurred, as in the figure.  This generates a corner in
$r(\theta,t)$ (though in the pictured example, the contour began with
corners at the maxima).  

There are also problematic points around a local minimum of $r$.  
Let $(r_{min},\theta_{min})$ be
the coordinates of the local minimum on the initial contour.  Since
$r_{min}>0$, all points $\theta>\theta_{min}$ ahead of this minimum will 
have traveled a nonzero angular distance along their characteristics at any 
$t>0$, which results in a growing gap of undefined points (i.e.~points
such that Eq.~(\ref{eqn:char}) has no solution in $\theta_0$ at
a given $t$) between $\theta_{min}$ and 
$\theta_{min}+Cr_{min}t$.  The obvious thing to do is to interpolate
by setting $r=r_{min}$ for all $\theta$ in this interval, as this is the only
natural way to ensure that the shape remains continuous.  
Thus intervals of constant $r$ (circular arcs) are
continually growing at local minima.

The two cases just described are the simplest cases where the
evolution along characteristics must be repaired to become continuous.  
There are several more similar
cases involving intervals of constant $r$
which lead to multivaluedness (or no-valuedness), but they are all
treated by either amputation or interpolation, as described above. 
In the terminology of Melikyan \cite{Melik98}, the points of amputation
are ``equivocal'' and the points of interpolation are
``dispersal''.

\section{Piecewise solution for $r(\theta,t)$}

From the considerations above, giving an explicit 
formula for the solution would involve several layers of if-then 
constructs.  We describe the full piecewise solution $r(\theta,t)$ to
Eq.~(\ref{eqn:ev0}) instead by decomposing the contour into
strictly monotonic intervals, and within each of these the solution 
depends continuously on the initial contour.  We also give
the positions of the endpoints separating
these intervals as a function of time.  

\begin{figure}
\centering
\includegraphics[width=7cm]{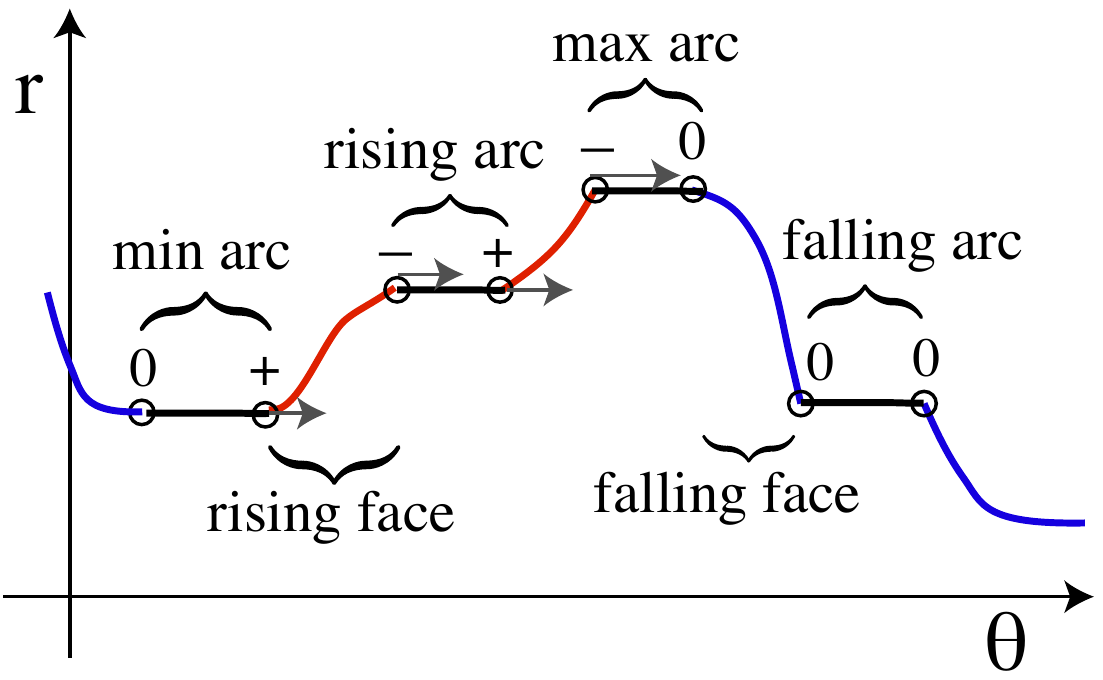}
\caption{(Color online) Decomposition of the contour into faces and
arcs; see text. The different types of endpoints are labeled with 
symbols: ``$0$'' are stationary, ``$+$'' are interpolating, ``$-$'' 
are amputating.}
\label{fig:decomposition}
\end{figure}

We define 
\textit{rising} and \textit{falling faces} of the contour to be 
connected components of points on the contour with $\partial_\theta
r>0$ and $\partial_\theta r<0$, respectively.  See
Fig.~\ref{fig:decomposition} for illustrations.  The step function in
Eq.~(\ref{eqn:char}) forces falling faces to be pieces of the original 
contour, i.e. $r(\theta,t)=r_0(\theta)$.  
The rising faces will be intervals from the original contour
``sheared'' by the evolution along characteristics.  More precisely,
$r(\theta,t)$ is defined implicitly by solving for the value of 
$\theta_0$ in Eq.~(\ref{eqn:char}) such that 
$\theta=\theta(\theta_0,t)$, and then setting 
$r(\theta,t)=r_0(\theta_0)$.  The shearing is caused by points moving 
with speed proportional to their radius.

We will call intervals of constant $r$ (circular) \textit{arcs}, and we will  
classify these into four types.  \textit{Rising} and \textit{falling arcs} 
are those that are adjacent to rising, respectively falling faces on both
sides.  \textit{Min} and \textit{max arcs} are those
which contain local minima, respectively maxima of the contour. 
Thus, $r(\theta,t)$ at any fixed time decomposes
into a set of faces and arcs. 
The evolution of each face or arc can be treated 
independently of the others for almost all times except for a 
discrete set of events when a face or arc changes into another type or
disappears.  

The endpoints of faces or arcs fall into three 
categories named according to their behavior under time 
evolution: \textit{stationary 
endpoints}, \textit{interpolating
endpoints}, and \textit{amputating endpoints}.  
Stationary endpoints are those that do not move under time evolution.  
There are two types, those at the right of a falling face and at the
left of a min or falling arc, and those at the right of a max or
falling arc and at the left of a falling face.  
Interpolating endpoints are those that move to the
right under time evolution and are the sites of new interpolation.
These are always to the right of rising or min arcs and to the left of
rising faces.  These endpoints move at constant speed $Cr$ where $r$
is the radius of the arc.  Finally, amputating endpoints move to the 
right and are the sites of new amputation.  They are always to the right
of rising faces and are on the left of max arcs, rising arcs, or
falling faces.  

To calculate the motion of the amputating endpoints we will need the 
survival time of each point of the pebble.  The intersection of the area under the 
contour with a circle of radius $r$ will be several disjoint arcs.  From
Eq.~(\ref{eqn:char}) these shrink in time as the left
endpoint of each moves with speed $Cr$ towards the right, until the
moment this value of $r$ is amputated and the entire interval has 
vanished.  The remaining lifetime of a point $(r,\theta)$ inside the
contour at time $t$ is thus $\frac{\Delta(r,\theta,t)}{Cr}$
where $\Delta(r,\theta,t)$ is the angular distance along the circle of
radius $r$ from $(r,\theta)$ to
the left endpoint of its interval on the contour at time $t$.  At fixed $r$, 
$\Delta(r,\theta,t)=\theta-\theta_L-Crt$, where $\theta_L$ is the 
position of this left endpoint  on the initial contour, i.e. $\theta_L$
is an appropriate solution of $r_0(\theta_L)=r$.

\begin{figure}
\centering
\includegraphics[width=7cm]{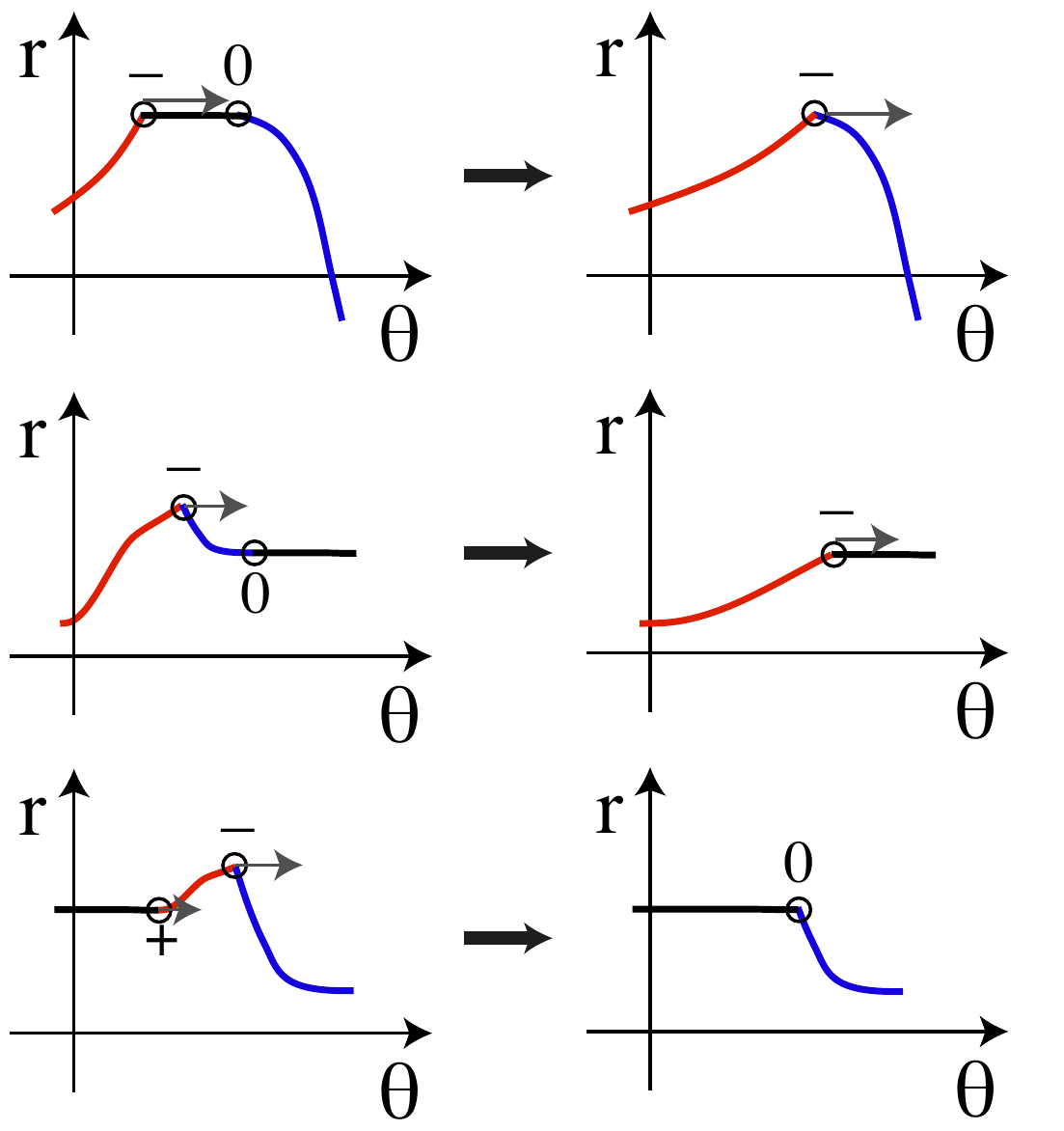}
\caption{(Color online) The face and arc merging processes, from top to
bottom: vanishing of a max arc, vanishing of a falling face, and
vanishing of a rising face. Endpoints are labeled 
as in Fig.~\ref{fig:decomposition}}
\label{fig:merging}
\end{figure}

For an amputating endpoint that lies between a rising and falling face,
the angular position as a function of time follows the contour of the
falling face. Its position is a solution $(r(t),\theta(t))$ of 
$\Delta(r,\theta,t)=Crt$.  One limit
to keep in mind is when the falling face is vertical. Then the
angular position of the amputating endpoint will be stationary 
for a period of time.  The opposite limit is when the amputating 
endpoint lies between a rising face and an arc with radius $r$, then the 
endpoint moves to the right with constant speed $Cr$.  
The nature of the above decomposition of the contour into faces and 
arcs changes precisely when endpoints collide with each other and
faces and arcs merge.  There are three basic processes: vanishing of a
max arc, vanishing of a falling face, and vanishing of a rising face.
See Fig.~\ref{fig:merging} for illustrations of these.  If a rising
face comes directly before a falling face, the radius $r_r$ of the arc
before the rising face and the radius $r_f$ of the arc after the
falling face determines which of the two faces vanish.  When
$r_r<r_f$, the falling face vanishes, when $r_r>r_f$ the rising face
vanishes, and when $r_r=r_f$, they vanish simultaneously and the two
arcs are joined.

To make the above solution a bit more concrete, we sketch what 
happens in the case of a rectangular contour with side lengths 
$2a$ and $2b$ with $a\leq b$.  Let $\gamma=\arctan(b/a)$.  The initial 
contour (for $0\leq\theta<2\pi$) takes the form
\begin{equation}
r_0(\theta)=\begin{cases}
a\sec\theta & 0\leq\theta<\gamma\text{ or
}2\pi-\gamma\leq\theta<2\pi\\
b\csc\theta & \gamma\leq\theta<\pi-\gamma\\
-a\sec\theta & \pi-\gamma\leq\theta<\pi+\gamma\\
-b\csc\theta & \pi+\gamma\leq\theta<2\pi-\gamma.\\
\end{cases}
\end{equation}

In the first instant of time, the local minima 
at $\theta=0,\pi/2,\pi,3\pi/2$ expand by interpolation into min 
arcs, so the initial contour consists of a min arc of zero width at 
each of these local minima, each sandwiched between a falling face and
a rising face.  Fig.~\ref{fig:frictionev} depicts the first interval
of time in the case $b/a=1$, where the amputating endpoints at the
local maxima (initially at positions
$\theta=\gamma,\pi\pm\gamma,2\pi-\gamma$) move to the right and the
interpolating endpoints at the local minima do as well.  
The solution in this first time interval consists 12 piecewise smooth 
curves, four each of min arcs, rising faces, and falling faces (though
there is a $\pi$ periodicity of the contour in $\theta$ 
which is preserved by the evolution, simplifying matters somewhat).
This proceeds until time $\tau_1=\frac{\pi/2-\arcsec(b/a)}{Cb}$ 
when the 
points on the contour above $r=b$ have all been amputated, and we 
remove the corresponding rising and falling faces.  If $a\neq b$, the
contour now consists of 8 piecewise smooth curves, two each of max arcs, 
min arcs, rising faces, and falling faces, as in the bottom left of 
Fig.~\ref{fig:tree}.  The
max arcs vanish at time $\tau_2=\frac{\arcsec(-b/a)-\arcsec(b/a)}{Cb}$.  
During the final phase of the evolution, the contour consists of 6 
piecewise smooth curves, two each of min arcs, rising faces and falling 
faces.  The rising faces and falling faces all vanish at time
$\tau_3=\frac{\pi}{Ca}$ and for all later times the contour is a 
circle with radius $r=a$.

\section{Coarse evolution and tree model}

The constructions in the previous section are a bit unwieldy to write
out by hand, though it is straightforward to program a computer to map 
$(\theta,t)$ to $r_0(\theta_0)$ and thus solve the evolution to the
precision of the initial data $r_0(\theta_0)$.  The complexity is all in
how the pattern of arcs and faces changes over time.  In this section,
we extract from the solution above a more combinatorial model of the 
evolution that focuses on this pattern and may make it more intuitive.

\begin{figure}
\centering
\includegraphics[width=7cm]{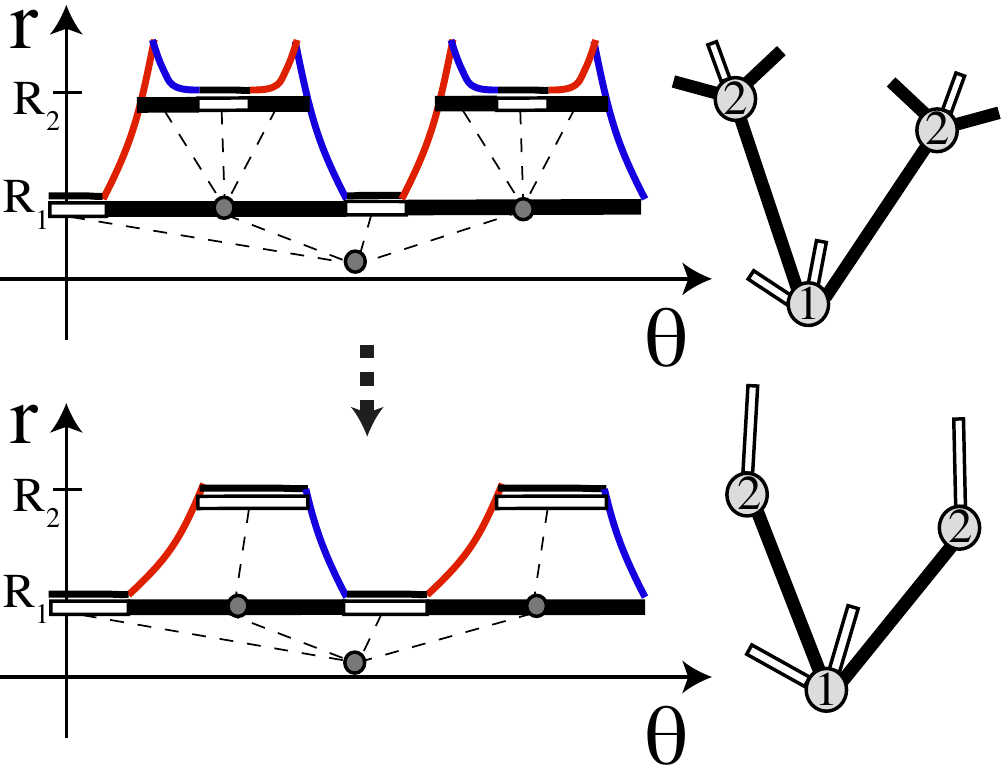}
\caption{(Color online) (top) A schematic contour for a rectangular
contour after some abrasion and its associated tree.  The two levels 
correspond to the distance $R_1$ from the origin to the long sides and the
distance $R_2$ to the short sides. The lengths of the edges,
corresponding to the widths of radial slices of the pebble, change in 
time according to their color and position -- black edges at level
$i$ shrink with rate $CR_i$, white edges sandwiched between
black edges grow with rate $CR_i$. (bottom) The contour and tree after
the four original corners of the rectangle have been amputated.}
\label{fig:tree}
\end{figure}

First, identify all values of $r$ such that the initial contour has a
point where $\partial_\theta r=0$ (do not count corners at maxima).  
Order these values from minimum to maximum to define 
$R_1<R_2<\cdots<R_N$.  Geometrically, these critical values are 
the radii of circles centered at the origin which are tangent to the 
contour.  These will be the $N$ ``levels'' of a
planar tree graph which we are constructing, which represents something 
like a skeleton of the contour.  For example, regular polygons have 
only one level, the distance from the origin to any edge, and the
rectangle discussed earlier has two levels, $a$ and $b$.  

Consider the set of intervals arising from the intersection of 
a circle of radius $R_i$ with the area below the contour.  Cut each
interval in this set into black and white edges as follows: every
subinterval which contains arcs (i.e. a segment on the contour 
coincident with the circle of radius $R_i$) becomes a white edge, and 
all other subintervals become black edges.  Note that the black edges 
correspond to subintervals which support some hump of the contour.  
Each edge is assigned a length equal to the angular width of its 
subinterval.  The edges just constructed constitute all the edges at
level $i$.  An example of white and black edge assignments can be 
seen on the left of Fig.~\ref{fig:tree}.

We will now glue these edges into a planar tree.
First, create a root vertex at level 1.  Attach one end of every edge at 
level 1 to this root vertex, preserving the cyclic ordering of edges. 
Next, create a vertex at every black edge whose corresponding hump has 
intervals at level 2 above it, and then attach one
end of every edge at level 2 to the appropriate vertex, preserving the
linear ordering.  Repeat this process of creating vertices and gluing
for each remaining level.  See Fig.~\ref{fig:tree} for the example of
a rectangle.  Roughly speaking, the tree captures the pattern of the
protrusions of the contour as we move from the origin outwards.  
Note that the edges at level $i$ are not necessarily all attached to 
edges at level $i-1$.

The dynamics of endpoints, faces and arcs yields the following rules
for the evolution of the tree.  As time progresses, every 
black edge at level $i$ shrinks at the rate $CR_i$.  White 
edges at level $i$ which happen to be sandwiched between black edges 
(in the cyclic ordering around the root if $i=1$, or the linear ordering 
above a vertex if $i>1$) grow at the 
same rate $CR_i$.  White edges at level $i>1$ which sit alone on a 
vertex shrink at the rate $CR_i$.  The lengths of all other white
edges are held constant.  If the length of an edge
shrinks to zero, we remove it; if two white edges become
adjacent on a vertex, we merge them into one white edge with length 
equal to the sum of their lengths.

Under this evolution, the tree contracts from the leaves
inwards; the edges supporting a branch will never vanish
before the edges at higher levels connected to it.  The total time of
evolution is thus determined by the length of the longest black edge
at level 1.  Translating back to the original contour, this means we
just need to measure the angular width of the base of the largest
hump.  Therefore the contours with fixed minimum radius which take 
longest to evolve to a circle are those with a single minimum radius.  
The tree picture also makes it clear that all contours evolve
to a circle with radius equal to the minimum radius of the initial
shape.  

At all times the pattern of white edges and black edges at different
levels on the tree may be used to reconstruct a coarse version of the
contour at that point in time.  This is not a one-to-one
correspondence between trees and contours, as there many possible contours
that lead to the same tree.  More explicitly, one can place at the
left and right of the intervals corresponding to each black edge an 
arbitrary increasing function (rising face) between the radii $R_i$ to 
$R_{i+1}$, respectively decreasing function (falling face), provided the 
angular widths of these two functions is consistent with the 
lengths of the edges above and the length of this black edge.  
Indeed, not even the maximum height of each hump enters this 
description.  However, all contours that lead to the same tree have
the same pattern of face and arc disappearances.  This property gives
some stability to the evolution -- if noise is added to the initial
contour, this will only affect the long term behavior insofar as it
might change the widths of the bases of the large scale features.
Small humps coming from the short-wavelength part of the noise will
correspond to short edges on the tree which quickly vanish or merge
with the large edges.

\section{Discussion}

The solution in this paper generalizes easily to the case where the 
equation takes the form $\partial_t r+f(r)\partial_\theta 
rH(\partial_\theta r)=0$ 
with nondecreasing $f$ in place of $Cr$.  If $f$ is not nondecreasing the
solution above will be modified significantly as
then some points of the contour would propagate in the opposite
direction. However, a
choice like this would also seem physically unmotivated.  The 
function $f$ allows for 
more general radius-speed relations, and the tree picture makes it 
clear that all that changes is the relative rate of growth or 
shrinkage of each edge, and not the overall qualitative picture; in 
particular, this may explain the observation in \cite{RMD10} that 
the model was robust to changing $r$ to $r^\alpha$ with 
$\alpha=1/2,1,2,3$.

We speculate next on some possible choices for $f$.  The rectangle 
is the most interesting case studied by Roth, Marques and
Durian, as it is the only contour with two widely-separated levels.
In their data (see rightmost panel of Fig.~2 in \cite{RMD10}), two 
of the corners are abraded before the other two,
whereas the model predicts that all four corners vanish at the same 
time ($\tau_2$ in the notation of the end of Sec.~III).  Thus 
the ``constant'' $C$ may differ between the rising 
faces, perhaps being larger if the radii of the preceding minimum is
smaller, and more generally, the speed $f$ might in general be a 
function not just of $r$ but also of $r_{min}$ of the face as well.  
In principle, (different branches of) $f(r)$ can be extracted from
experimentally measured curve contours by computing $\partial_t r$,
$\partial_\theta r$ and comparing them at fixed $r$, but preliminary 
analysis of data provided by Roth, Marques and Durian was not conclusive 
due to the difficulty of estimating $\partial_t r$ accurately.

We did not carry out an
extensive comparison here with the numerical solution of \cite{RMD10}, but
the plotted curves appeared indistinguishable in a few checks.
Indeed, it would be interesting to place this work on firmer
mathematical ground along the lines of \cite{Melik98} by analyzing 
how the corners generated by the
amputation and interpolation processes are smoothed by the addition of
higher derivative terms, and how this happens in the finite difference
scheme of Roth, Marques and Durian.  We also did not yet attempt to
calculate the typical evolution of area, perimeter and other geometric 
quantities from our exact solution.  Finally, we leave open the 
question whether erosion or shape evolution models in general may be 
simplified by posing them as laws
for evolving tree graphs.  In particular, it may be easier to
construct models for the evolution of an ensemble of pebbles in terms of 
a mean-field model on trees, rather than attempt a more direct
description of ensembles of interacting contours.  

\begin{acknowledgments}
I thank A.~Roth and D.~Durian for helpful discussions 
and for generously sharing experimental data and G.~Alexander for
useful feedback on a draft.  The author is grateful for the continued 
encouragement of R.~Kamien as well as support from NSF Grant No.~DMR05-47230.
\end{acknowledgments}

\end{document}